# MHD flow and heat transfer due to the axisymmetric stretching of a sheet with induced magnetic field

Tarek M. A. El-Mistikawy

Department of Engineering Mathematics and Physics, Faculty of Engineering, Cairo University, Giza 12211, Egypt

**Abstract**

The full MHD equations, governing the flow due to the axisymmetric stretching of a sheet in the presence of a transverse magnetic field, can be cast in a self similar form. This allows evaluation of the induced magnetic field and its effect on the flow and heat transfer. The problem involves three parameters- the magnetic Prandtl number, the magnetic interaction number, and the Prandtl number. Numerical solutions are obtained for the velocity field, the magnetic field, and the temperature, at different values of the magnetic Prandtl number and the magnetic interaction number. The contributions of the viscous dissipation, Joule heating, and streamwise diffusion to the heat flux toward the sheet are assessed.

## 1      Introduction

Flow due to the axisymmetric stretching of a sheet was first introduced by Wang [1] as a special case of the three dimensional flow due to the stretching of a flat surface. Its practical application is to extrusion processes and polymer and glass industries

The problem allowed self similarity of the governing Navier-Stokes equations. This encouraged researchers to add new features that maintained self similarity. Ariel [2] and Hayat et al. [3] added partial slip. Ariel [4] added suction. Ariel [5,6] and Hayat et al. [7] considered non-Newtonian fluids. The latter article included heat transfer with improper transformation of the temperature, violating self similarity. The magnetohydrodynamic (MHD) flow of conducting fluids was handled by Ariel et al. [4] and Ariel [8]. Sahoo [9] treated the MHD flow of second grade fluids. The assumption of small magnetic Reynolds number was adopted, leaving the imposed transverse magnetic field unaltered by induction.

Self similar formulation of flow problems is of such considerable value. Reduced to ordinary differential equations, the problems are amenable to different analysis and solution methods, and allow for evaluation and comparison of these approaches. For the current problem, numerical solutions adopting the shooting technique were presented by Wang [1] and Fang [10]. Variants of the homotopy perturbation method were used by Ariel [2,11] and Ariel et al. [4]. Hayat and co-workers [3,7] implemented the homotopy analysis method. Ackroyd's method [12], as well as a modification of which, was used by Ariel [2,6,8,13]. Ariel also used residual minimization [6,8,13]. Perturbation expansions were developed for small and large slip coefficient [2] and second grade fluid parameter [6]. Comparisons of the methods are found in [2,3,6,8].

In this article, the full MHD flow equations; namely, the continuity, momentum, energy, and Maxwell's equations are shown to admit self similar transformation. Numerical solutions are obtained for the velocity, induced magnetic field, and temperature. The effect of the induced magnetic field on the flow and heat transfer is demonstrated. Traditionally ignored heat generation and transfer processes such as viscous dissipation, Joule heating and streamwise diffusion are assessed.

## 2  Mathematical model

An electrically conducting incompressible Newtonian fluid is driven by the axisymmetric stretching of a non-conducting sheet. The stretching speed along the radial $r$-direction is $\omega r$, where the stretching rate $\omega$ is constant. In the farfield as the transverse coordinate $z \sim \infty$, the fluid is essentially quiescent under pressure $p_\infty$ and temperature $T_\infty$, and is permeated by a stationary magnetic field in the $z$-direction of strength $B$.

In the absence of an imposed and induced electric field, the equations governing this steady MHD flow are

$$u_r + \frac{u}{r} + w_z = 0 \tag{1}$$

$$\rho(uu_r + wu_z) + p_r = \rho\nu(u_{rr} + \frac{u_r}{r} - \frac{u}{r^2} + u_{zz}) - \sigma(su - qw)s \tag{2}$$

$$\rho(uw_r + ww_z) + p_z = \rho\nu(w_{rr} + \frac{w_r}{r} + w_{zz}) + \sigma(su - qw)q \tag{3}$$

$$s_r - q_z = \sigma\mu(su - qw) \tag{4}$$

$$q_r + \frac{q}{r} + s_z = 0 \tag{5}$$

$$\rho c(uT_r + wT_z) = k(T_{rr} + \frac{T_r}{r} + T_{zz}) + \rho\nu[2(u_r^2 + \frac{u^2}{r^2} + w_z^2) + (u_z + w_r)^2] + \sigma(su - qw)^2 \tag{6}$$

where $(u, w)$ are components of the velocity and $(q, s)$ are components of the magnetic field in the $(r, z)$ directions, respectively, $p$ is the pressure and $T$ is the temperature. Constant are the fluid density $\rho$, kinematic viscosity $\nu$, electric conductivity $\sigma$, magnetic permeability $\mu$, specific heat $c$, and thermal conductivity $k$.

Our interest being the evaluation of the magnetic field and its effect on the flow and heat transfer, we opt to invoke the simple surface conditions of no-slip, non-porosity, and freestream temperature.

$$z = 0: \quad u = \omega r, \quad w = 0, \quad T = T_\infty \tag{7}$$

The farfield conditions are

$$z \sim \infty: \quad u \sim 0, \quad p \sim p_\infty, \quad T \sim T_\infty, \quad q \sim 0, \quad s \sim B \tag{8}$$

the last two of which are consistent with the physical requirement of the farfield being free from any current density.

The problem admits the similarity transformations:

$$z = (\nu/\omega)^{1/2}\zeta, \quad w = -2(\nu\omega)^{1/2} f(\zeta), \quad u = \omega r f' \tag{9}$$

$$s = B[1 + 2\sigma\mu\nu g(\zeta)], \quad q = -B\sigma\mu(\nu\omega)^{1/2} r g' \tag{10}$$

$$p = p_\infty - 2\rho\nu\omega[f' + f^2 - f^2(\infty)] - \tfrac{1}{2} B^2\sigma^2\mu\nu\omega r^2 g'^2 \tag{11}$$

$$T = T_\infty + (\nu\omega/c)[\phi(\zeta) + (\omega/\nu) r^2 \theta(\zeta)] \tag{12}$$

where primes denote differentiation with respect to $\zeta$. Expression (11) for the pressure indicates radial-wise variation, due to the induced magnetic field.

The problem becomes

$$f''' + 2ff'' - f'^2 - \beta f' = P_m \beta [4 g f'(1 + P_m g) - 2 g' f(1 + 2P_m g) - g'^2] \tag{13a}$$

$$f(0) = 0, \quad f'(0) = 1, \quad f'(\infty) = 0 \tag{13b}$$

$$g'' = f' + 2P_m (gf' - fg') \tag{14a}$$

$$g(\infty) = 0, \quad g'(\infty) = 0 \tag{14b}$$

$$\frac{1}{P_r}\theta'' + 2f\theta' - 2f'\theta = -\beta g''^2 - f''^2 \tag{15a}$$

$$\theta(0) = 0, \quad \theta(\infty) = 0 \tag{15b}$$

$$\frac{1}{P_r}\phi'' + 2f\phi' = -4\frac{1}{P_r}\theta - 12 f'^2 \tag{16a}$$

$$\phi(0) = 0, \quad \phi(\infty) = 0 \tag{16b}$$

where $P_m = \sigma\mu\nu$ is the magnetic Prandtl number, $\beta = \sigma B^2/\rho\omega$ is the magnetic interaction number, and $P_r = \rho\nu c/k$ is the Prandtl number. Note that the present choice of characteristic length $(\nu/\omega)^{1/2}$ and velocity $(\nu\omega)^{1/2}$ renders $P_m = R_m$, the magnetic Reynolds number.

For practical applications, $P_m$ is much smaller than unity [14]. For negligible $P_m$, the velocity field is uncoupled from the magnetic field and is governed by the following problem

$$f''' + 2ff'' - f'^2 - \beta f' = 0, \; f(0) = 0, \; f'(0) = 1, \; f'(\infty) = 0 \tag{17}$$

This is the same problem formulated and solved by Ariel [8], under the assumption of negligible $R_m$. Numerical results for $f(\zeta)$, $f'(\zeta)$, $f''(0)$, and $f(\infty)$ for different values of $\beta$ are found therein.

The corresponding problem for $g$ is

$$g'' = f', \; g(\infty) = 0, \; g'(\infty) = 0 \tag{18}$$

with the solution

$$g' = f(\zeta) - f(\infty), \; g = \int_\zeta^\infty [f(\infty) - f(\eta)] d\eta \tag{19}$$

so that

$$g'(0) = -f(\infty), \; g(0) = \int_0^\infty [f(\infty) - f(\eta)] d\eta \tag{20}$$

The value of $f(\infty)$ is obtained from the solution of the problem (17) for $f$.

## 3      Numerical method

Since a closed form solution is not possible, we seek an iterative numerical solution. In the $n^{th}$ iteration, we solve, for $f_n(\eta)$, problem (13a,b) with the right hand side of Eq. (13a) evaluated using the previous iteration solutions $f_{n-1}(\eta)$ and $g_{n-1}(\eta)$. Then we solve, for $g_n(\eta)$, Eq. (14a) with the known $f_n(\eta)$, together with conditions (14b). The iterations continue until the maximum error in $f(\eta_\infty)$, $f''(0)$, $g(0)$ and $g'(0)$ becomes less than a prescribed tolerance $\varepsilon = 10^{-10}$. For the first iteration, we zero the right hand side of Eq. (13a) which corresponds to $g_0(\eta) = 0$.

The numerical solution of the problems for $f_n(\eta)$ and $g_n(\eta)$ utilizes Keller's two point, second order accurate, finite-difference scheme [15]. A uniform step size $\Delta \eta$ is used on a finite domain $0 \leq \eta \leq \eta_\infty$. The value of $\eta_\infty$ is chosen sufficiently large in order to insure the asymptotic satisfaction of the farfield conditions. The non-linear terms in the problem for $f_n(\eta)$ are quasi-linearized, and an iterative procedure is implemented; terminating when the maximum error in $f_n(\eta_\infty)$ and $f_n''(0)$ becomes less than $\varepsilon$.

Having determined $f(\eta)$ and $g(\eta)$, we solve the linear problem (15a,b) for $\theta(\eta)$, then (16a,b) for $\phi(\eta)$, using Keller's scheme.

## 4   Results and discussion

The results presented below are intended to explore the effect of the induced magnetic field on the flow and heat transfer. Of interest are the surface shear, the entrainment rate, the $r$ and $z$ components of the induced magnetic field at the surface, and the constant and radial-wise varying constituents of the heat flux at the surface, which are represented by $f''(0)$, $f(\eta_\infty)$, $g(0)$, $g'(0)$, $\phi'(0)$ and $\theta'(0)$, respectively.

The problems for $f(\eta)$ and $g(\eta)$ involve two parameters, the magnetic Prandtl number $P_m$ and the magnetic interaction number $\beta$. The problems for $\theta(\eta)$ and $\phi(\eta)$ involve the Prandtl number $P_r$, as a third parameter. All results presented below are for $P_r = 0.72$.

Tables 1 and 2 demonstrate the effect of varying $P_m$ when $\beta = 1$, and varying $\beta$ when $P_m = 0.1$, respectively. As $P_m$ decreases, there is obvious tendency to the limiting case of $P_m = 0$. As $\beta$ increases, the fluid motion is restrained more and more, so that the induced components of the magnetic field decrease. The flow shapes as a boundary layer that diminishes in size leading to reduction in the rate of fluid entrainment, and rise in surface shear.

On the right-hand-sides of Eqs. (15a) and (16a), the first terms represent Joule heating and streamwise heat diffusion, respectively, while the second terms represent heat dissipation. Tables 3 and 4 demonstrate the effect of these three processes. The predicted heat flux to the surface, represented by $\phi'(0)$ and $\theta'(0)$, is the additive effect of viscous dissipation and Joule heating. When both effects are neglected, the thermal problem predicts zero flux. As expected, the larger the magnetic field the greater the contribution of Joule heating. As $\beta$ increases, the contribution of the viscous dissipation to $\theta'(0)$ increases, because of the rise in $f''^2$. The fall then rise of $\phi'(0)$ can be explained in view of the last two columns in Table 3, which dissect the viscous dissipation into its constituents due to $f'^2$ and $f''^2$. The higher the value of $|f''|$, the curvature of the $f$ profile, the faster the drop in the slope $f'$; hence, as $\beta$ increases, the effect of $f''^2$ rises while the effect of $f'^2$ falls. The streamwise diffusion manifests itself through $\phi'(0)$ relaying to it the Joule heating as well as the part of the viscous dissipation due to $f''^2$.

Profiles of the velocity and induced magnetic field components are depicted in Fig. 1, for the typical case of $P_m = 0.1$ and $\beta = 1$. Corresponding profiles of the temperature constituents are shown in Fig.2. It is noted that the first set of profiles (in Fig. 1) reach the farfield conditions much faster than the second set (in Fig. 2). As $\beta$ increases, this becomes more prominent. The farfield conditions are reached progressively faster by the first set and progressively slower by the second set.

## 5 Conclusion

The problem of the flow due to the axisymmetric stretching of a sheet in the presence of a transverse magnetic field has been shown to admit self similarity of the full MHD governing equations. Numerical solutions have been obtained, revealing samples of which have been demonstrated. The following conclusions are drawn.

The self similar formulation indicates radial variation in the pressure due to the induced magnetic field, and in the temperature even when the surface temperature is constant.

No surface conditions on the induced magnetic field should be imposed. Rather, the requirement of zero current density in the farfield should be honored. Note that the vectors of velocity $\mathbf{V}$ (of magnitude $-2\sqrt{\nu\omega}f(\infty)$) and magnetic field $\mathbf{B}$ are parallel, in the farfield; hence, the current density vector $\mathbf{J} = \sigma\mathbf{V}\times\mathbf{B}$ vanishes.

As the magnetic Prandtl number $P_m$ diminishes, the problem and its solution approach those with $P_m = 0$. This is in accord with the conclusion of El-Mistikawy [16] in the corresponding two-dimensional case.

The increase of the imposed magnetic field results in restraint of the flow, reducing the velocity and, consequently, the induced magnetic field. Close to the surface, Joule heating increases, while viscous dissipation decreases then increases; being comprised of two parts, one rising and one falling.

Streamwise diffusion is important. It relays information from one constituent of the temperature to the other one.

Finally, it is noted that features such as surface feed (suction or injection), velocity slip, thermal slip, and prescribed surface temperature or heat flux can be incorporated in the self-similar formulation.